\documentclass{ws-ijmpa}

\usepackage{amsmath}
\def\fsl#1{\setbox0=\hbox{$#1$}                 
   \dimen0=\wd0                                 
   \setbox1=\hbox{/} \dimen1=\wd1               
   \ifdim\dimen0>\dimen1                        
      \rlap{\hbox to \dimen0{\hfil/\hfil}}      
      #1                                        
   \else                                        
      \rlap{\hbox to \dimen1{\hfil$#1$\hfil}}   
      /                                         
   \fi}                                         %

\def\nnb{\nonumber}

\def\be{\begin{equation}}
\def\ee{\end{equation}}

\newcommand{\bea}{\begin{eqnarray}}
\newcommand{\eea}{\end{eqnarray}}
\newcommand{\bwt}{\begin{widetext}}
\newcommand{\ewt}{\end{widetext}}

\def\wsep{ \nnb \\ &&}

\def\eed{\end{document}}

\def\al{\alpha}

\def\al{\alpha}
\def\be{\beta}

\def\al{{\alpha}}
\def\my#1{ #1 }

\def\sbbkf#1{\bigg ( #1 \bigg )}

\def\m_z{m_{\textrm {Z}}}

\begin{document}

\markboth{Qi-Shu Yan}{Renormalization of the EWCL and its Application to LEP2}

\catchline{}{}{}{}{}

\title{Renormalization of the EWCL and its Application to LEP2}

\author{Qi-Shu Yan}
\address{Theory Group, KEK,  Tsukuba, 	 305-0801, Japan.}

\maketitle

\vspace{-5.6cm}

\hfill {\small TKYNT-05/21 }

\vspace{5.6cm}


\begin{abstract}

We perform a systematic one-loop renormalization on the electroweak chiral Lagrangian (EWCL) 
up to $O(p^4)$ operators and construct the renormalization 
group equations (RGE) for the anomalous couplings.
We examine the impact of the triple gauge coupling (TGC) measurement from LEP2 to the uncertainty
of the $S-T$ parameter at the $\Lambda=1 TeV$, and find that the
uncertainty in the TGC measurements can shift $S(\Lambda)$ at
least $3.3 \sigma$. 

\end{abstract}



\section{Introduction}
Direct precision measurement on the TGC at LEP2 reinforces  
\cite{Heister:2001qt,Abbiendi:2003mk,Achard:2004ji,Schael:2004tq,LEPEWWG,Abazov:2005ys} 
our belief that the standard model is the correct description for the particle physics phenomenology.
However, Higgs is still an elusive particle and the electroweak symmetry breaking
mechanism is still a mystery. Therefore, it is reasonable and necessary to  
interpret the Z pole data and LEP2 TGC data in a model
independent fashion without a Higgs boson, so as to give some guides to
model-building works. Such an effective description
is given as the EWCL 
\cite{Appelquist:1980vg,Longhitano:1980iz,Appelquist:1993ka}, where the Goldstone particles
are parameterized in nonlinear form.

In order to extrapolate electroweak precision data collected 
at low energy region $\mu=\m_z$ up to
the ultraviolet cutoff $\Lambda=1 TeV$, radiative corrections
of the EWCL must be correctly and efficiently summed over.
There are several groups who have considered the radiative corrections
within the EWCL by including anomalous couplings, 
due to different assumptions and technique uncertainties, 
those results do not agree with each other 
\cite{DeRujula:1991se,Hagiwara:1992eh,Hernandez:1993pp,Burgess:1993qk,Dawson:1994fa,vanderBij:1997ec}.
For example, the $\beta$ function of anomalous couplings in the unitary gauge with dimensional
regularization \cite{Burgess:1993qk,Dawson:1994fa} do not agree 
with well-known results \cite{Longhitano:1980iz}.
It seems a puzzle 
whether the radiative corrections in the EWCL with dimensionless
anomalous couplings is
well-defined or not \cite{DeRujula:1991se}.

Therefore our project is to perform such a 
systematic renormalization for the EWCL to investigate this puzzle.
We use the background field method, a renormalizable gauge,
path integral, dimensional and heat kernel regularization,
${\overline {MS}}$ renormalization scheme \cite{Yan:2002rt}, to perform the
systematic one-loop expansion to the EWCL and construct the
RGE of those dimensionless 
anomalous couplings.
There are some preliminary results \cite{try}, 
and now they are almost finished \cite{now}.

We use our RGE method to analyze 
the electroweak precision data at Z poles and TGC.

\section{The renormalization group equations}
We construct the counter terms
for all $O(p^2)$ and $O(p^4)$ operators at one 
loop level and succeed in building the
RGE. Our power counting rule in loop expansion is equivalent to the large $N$ 
and standard mass dimension
power counting rules. In the realistic
analysis, we can keep only linear terms of 
anomalous couplings in the $\beta$ function \cite{Georgi:1992dw}.
By assuming that ghosts are complex ghosts, we handle the nonhermitean
ghost term in a simple way.
Then, the $\beta$ functions of two point parameters are given as:
\bea
\beta_{\al_1}&=&
\frac{1}{12} + 2 \my{\al_1} g^2 + \my{\al_8} g^2  + \frac{5 \my{\al_2} g^2}{2} - 
  \frac{5 \my{\al_3} g^2}{6} + 
  \frac{\my{\al_9} g^2}{2} \label{betaa1}
\,,\\
\beta_{\al_8}&=&{\beta \over 2} - 
  \my{\al_1} {\my{g'}}^2+
\frac{37 \my{\al_8} g^2}{6} + 
  \frac{5 \my{\al_2} {\my{g'}}^2}{6} - 
  \frac{\my{\al_3} {\my{g'}}^2}{2}
+ \frac{17 \my{\al_9} g^2}{6}  
\,,\\
\beta_{\beta}&=&
-\frac{3 {\my{g'}}^2}{8} -{15 \beta g^2 \over 4} - {3 \beta {\my{g'}}^2\over 4 } + 
  \my{\al_1} \sbbkf{ \frac{9 g^2 {\my{g'}}^2}{4} + 
     \frac{{\my{g'}}^4 g^2}{2 G^2}  
}  + \my{\al_8} \sbbkf{ \frac{g^4}{8} -\frac{{\my{g'}}^2 g^4}{4 G^2}  } \wsep + 
  \my{\al_2} \sbbkf{ \frac{5 g^2 {\my{g'}}^2}{2} - 
     \frac{3 {\my{g'}}^4}{4} } + 
  \my{\al_3} \frac{5  g^2 {\my{g'}}^2}{2} + 
  \my{\al_9} \sbbkf{ -\frac{g^4}{2} + 
     \frac{3 g^2 {\my{g'}}^2}{4} }  \wsep - 
  \my{\al_4} \sbbkf{ \frac{15 g^2 {\my{g'}}^2}{4} + 
     \frac{15 {\my{g'}}^4}{8} }  - 
  \my{\al_5} \sbbkf{ \frac{3 g^2 {\my{g'}}^2}{2} +
     \frac{3 {\my{g'}}^4}{4} } \wsep - 
  \my{\al_6} \sbbkf{ \frac{3 g^4}{4} + 
     \frac{33 G^4}{8} }  - 
  \my{\al_7} \sbbkf{ 3 g^4 + 3 G^4 }    - 
  \my{\al_{10}} \sbbkf{ \frac{9 G^4}{2} }   
\,.
\label{rge1}
\eea
Here we have organized these $\beta$ functions 
in the order of the contributions
from the constant terms and the quadratic, triple, and quadruple anomalous couplings.

We omitted other $\beta$ functions here due to the reason that they are not
directly related with $S-T$ fitting.

\section{Application to precision data}
In order to be as general as possible, we have
not imposed the custodial symmetry in our analysis.

Although D0 collaboration at 
Tevatron \cite{Abazov:2005ys} has reported their measurements on the TGC,
errors are much larger than those at LEP2. Therefore we use the 
data published by the L3 collaboration and the combined fit results
from LEP EW working group \cite{LEPEWWG} in our analysis.

We have taken
$\delta  k_Z = -.076\pm 0.064$ from L3 \cite{Achard:2004ji},  and 
$\delta  k_\gamma =-.027 \pm 0.045 $ and $\delta  g_Z^1 = -0.016 \pm 0.022$
from \cite{LEPEWWG} as inputs to determine the anomalous couplings 
$\al_2$, $\al_3$, and $\al_9$.
These data are extracted from one-parameter TGC fits and the last two are extracted by
imposing the custodial symmetry. Each of data corresponds to 
a set of solution for $\al_2$ , $\al_3$, and $\al_9$.
In order to combine these data in our analysis, we assume that these data are extracted from
independent measurements (From the theoretical viewpoint, these measurements must be correlated
but the the correlation is energy-dependent).

With the above assumption, we find
\begin{equation}
\label{3pfit}
\begin{array}{rl}
\al_2 =  \!\!\!&(-0.09\pm 0.14)\\
\al_3 =  \!\!\!&(-0.03\pm 0.04)\\
\al_9 =  \!\!\!&( 0.12\pm 0.12)
\end{array}
\rho_{corr.} = \left(\begin{matrix}
1 \,\,\,& & \cr 
0 & 1 \,\,\,& \cr 
-.68&-.32&1\,\,\,\cr
\end{matrix}
\right).
\end{equation}

The $S(\Lambda)$ and $T(\Lambda)$  are depicted in Fig. 1.
The solid curves correspond to the analysis including the contributions of the 
TGC, while the dashed curves correspond to the analysis discarding 
the contributions of the TGC. 
The effects of ultraviolet cutoff are shown by 
taking three values of ultraviolet cutoff $\Lambda=$
 $300$ GeV, $1$ TeV, and $3$ TeV, respectively. 
The $S-T$ contour at $\mu=\m_z$ is depicted as a reference 
contour to compare. 
\begin{figure}
\begin{center}
\includegraphics[scale=0.35]{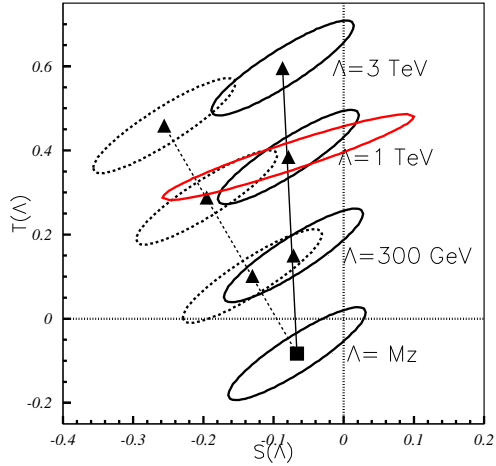}
\end{center}
\caption{ $S(\Lambda)-T(\Lambda)$ contours at $\Lambda=\m_z$, $300$ GeV,
$1$ TeV, and $3$ TeV, respectively. Solid curves have incorporated
the contributions of the TGC, while the dashed curves haven't.} 
\label{fig1}
\end{figure}

Without including the contribution of the TGC, $S(\Lambda)$ goes to the negative
value direction with the increase of $\Lambda$ in agreement with
the observation of Ref. \cite{Bagger}. 
However, when taking into account the contributions of the TGC with the
central value of LEP2 fit in Eq. (\ref{3pfit}), we  
observe that the allowed region of  $S(\Lambda)$ and $T(\Lambda)$ shifts toward
positive $S(\Lambda)$ and positive $T(\Lambda)$.  

The contour generated by including $1 \,\sigma $ errors  of 
$\al_2-\al_3-\al_9$ (the largest contour in 
Fig. 1 at $\Lambda=1$ TeV) shows 
that $S(\Lambda)$ can vary from $-0.26$ to $0.1$.
We observe that the uncertainty in the anomalous TGC can swing
the central value of $S(\Lambda)$ at least $3.3 \sigma$ away. 

\section{Conclusion}
We have performed the one-loop systematic 
renormalization on the EWCL. Our results
agree with well-known results \cite{Longhitano:1980iz} 
if anomalous couplings
vanish (sign differences are 
due to the Euclidean space convention
and the definition of covariant differential operator). 
Our results show that in the EWCL those dimensionless anomalous couplings
run in a logarithmic way. 
However, there is still one puzzle left: what's the meaning of nonhermitean
ghost term? What's the correct method to treat it?

We have analyzed the uncertainty of $S(\Lambda)$ caused by the uncertainty of
TGC measurements, which can swing the $S(\Lambda)$
at least $3.3 \sigma$ away.

\section*{ Acknowledgements}
The author would like to thank Prof. T. Appelquist, Prof. Y.P. Kuang, Prof. M. Tanabashi, 
Prof. Q. Wang, and Prof. H. Q. Zheng for helpful and stimulating discussions.
Especial thanks are indebted to Prof. K. Hagiwara for constant encouragements during this project.
The project is supported by the 
Japan Society for the Promotion of Science (JSPS) fellowship program.

\end{document}